\documentclass[runningheads,a4paper]{llncs}

\usepackage{amssymb}
\usepackage{graphicx}

\usepackage{url}
\urldef{\mailsa}\path|{binh, takasu}@nii.ac.jp|
\newcommand{\keywords}[1]{\par\addvspace\baselineskip
\noindent\keywordname\enspace\ignorespaces#1}

\usepackage{amsmath}
\usepackage{multirow}
\usepackage{upgreek}
\usepackage{bm}

\usepackage{makecell}

\begin{document}
\mainmatter

\title{A Probabilistic Model for the Cold-Start Problem in Rating Prediction using Click Data}
\author{ThaiBinh Nguyen\inst{1}
\and Atsuhiro Takasu\inst{1,2}}
\titlerunning{A Probabilistic Model for Rating Prediction using Click Data}
\authorrunning{ThaiBinh Nguyen, Atsuhiro Takasu}
\institute{Department of Informatics,\\SOKENDAI (The Graduate University for Advanced Studies), Tokyo, Japan\\
\and
National Institute of Informatics, Tokyo, Japan\\
\mailsa\\}

\maketitle

\begin{abstract}
One of the most efficient methods in collaborative filtering is matrix factorization, which finds the latent vector representations of users and items based on the ratings of users to items. However, a matrix factorization based algorithm suffers from the \textit{cold-start} problem: it cannot find latent vectors for items to which previous ratings are not available. This paper utilizes \textit{click data}, which can be collected in abundance, to address the cold-start problem. We propose a \textit{probabilistic item embedding} model that learns item representations from click data, and a model named EMB-MF, that connects it with a \textit{probabilistic matrix factorization} for rating prediction. The experiments on three real-world datasets demonstrate that the proposed model is not only effective in recommending items with no previous ratings, but also outperforms competing methods, especially when the data is very sparse.
\end{abstract}

\section{Introduction}
Rating prediction is one of the key tasks in recommender systems. From research done previously on this problem, matrix factorization (MF) \cite{salakhutdinov2008a,koren2008factorization,Gopalan:2015:SRH:3020847.3020882} is found to be one of the most efficient techniques. An MF-based algorithm finds the \textit{vector representations} (latent feature vectors) of users and items, and uses these vectors to predict the unseen ratings. However, an MF-based algorithm suffers from the \textit{cold-start} problem: it cannot find the latent feature vectors for items that do not have any prior ratings; thus cannot recommend them.

To address the \textit{cold-start} problem, many methods have been proposed. Most of them rely on exploiting \textit{side information} (e.g., the contents of items). The collaborative topic model \cite{wang2011collaborative} and content-based Poisson factorization \cite{conf_nips_GopalanCB14} use text content information of items as side information for recommending new items. In \cite{conf_kdd_WangWY15}, the author proposed a model for music content using deep neural network and combined it with an MF-based model for music recommendation. However, in many cases, such side information is not available, or is not informative enough (e.g., when an item is described only by some keywords).

This paper focuses on utilizing \textit{click data}, another kind of feedback from the user. The advantage of utilizing click data is that it can be easily collected with abundance during the interactions of users with the systems. The idea is to identify item representations from click data and use them for rating prediction.

The main contributions of this work can be summarized as follows:
\begin{itemize}
  \item We propose a \textit{probabilistic item embedding} model for learning item embedding from click data. We will show that this model is equivalent to performing PMF \cite{salakhutdinov2008a} of the positive-PMI (PPMI) matrix, which can be done efficiently.
  \item We propose EMB-MF, a model that combines the \textit{probabilistic item embedding} and PMF \cite{salakhutdinov2008a} for coupling the item representations of the two models.
  \item The proposed model (EMB-MF) can automatically control the contributions of prior ratings and clicks in rating predictions. For items that have few or no prior ratings, the predictions mainly rely on click data. In contrast, for items with many prior ratings, the predictions mainly rely on ratings.
\end{itemize}

\section{Proposed Method}
\label{sec:methodology}
\subsection{Notations}
The notations used in the proposed model are shown in Table \ref{tab:notations}.
\begin{table}[ht]
  \caption{Definitions of some notations}
  \label{tab:notations}
  \centering
  \renewcommand{\arraystretch}{1.2}
  \begin{tabular}{c|l}
    \hline
    Notation&Meaning\\
    \hline
    $N, M$ & the number of users, number of items\\
    $R, S$ & the rating matrix, PPMI matrix\\
    $\mathcal{R}$ & the set of $(u,i)$-pair that rating is observed (i.e., $\mathcal{R}=\{(u,j)|R_{uj}>0\}$\\
    $\mathcal{R}_u, \mathcal{R}_i$ & the set of items that user $u$ rated, set of users that rated item $i$\\
    $\mathcal{S}$ & the set of $(i,j)$-pair that $S_{ij}>0$ (i.e., $\mathcal{S}=\{(i,j)|S_{ij}>0\}$\\
    $\mathcal{S}_i$ & the set of item $j$ that $S_{ij}>0$ (i.e., $\mathcal{S}_i=\{j|S_{ij}>0\}$\\
    $d$ & the dimensionality of the feature space\\
    $\mathbf{I}_d$ & the $d$-dimensional identity matrix\\
    $\rho_i$, $\alpha_i$ & the \textit{item embedding vector} and \textit{item context vector} of item $i$\\
    $\theta_u$, $\beta_i$ & the \textit{feature vector} of user $u$, \textit{feature vector} of item $i$, in the rating model\\
    $\mu$, $b_u$, $c_i$ & the \textit{global mean} of ratings, \textit{user bias} of user $u$, \textit{item bias} of item $i$\\
    $B_{ui}$ & $b_u + c_i + \mu$\\
    $\bm{\rho}, \bm{\alpha}, \bm{\theta}, \bm{\beta}, \mathbf{b}, \mathbf{c}$ & $\{\rho\}_{i=1}^{M}$, $\{\alpha\}_{i=1}^{M}$, $\{\theta\}_{u=1}^{N}$, $\{\beta\}_{i=1}^{M}$, $\{b\}_{u=1}^{N}$, $\{c\}_{i=1}^{M}$\\
    $\bm{\Omega}$ & the set of all model parameters (i.e., $\bm{\Omega}=\{\bm{\rho}, \bm{\alpha}, \bm{\theta}, \bm{\beta}, \mathbf{b}, \mathbf{c}\}$)\\
  \hline
\end{tabular}
\end{table}
\subsection{Probabilistic Item Embedding Based on Click Data}
The motivation behind the use of clicks for learning representations of items is the following: if two items are often clicked in the \textit{context} of each other, they are likely to be similar in their nature. Therefore, analyzing the \textit{co-click} information of items can reveal the relationship between items that are often clicked together.

\textit{``Context"} is a modeling choice and can be defined in different ways. For example, the context can be defined as the set of items that are clicked by the user (\textit{user-based} context); or can be defined as the items that are clicked in a session (\textit{session-based} context). Although we use the user-based context to describe the proposed model in this work, other definitions can also be used.

We represent the association between items $i$ and $j$ via a \textit{link function} $g(.)$, which reflects how strong $i$ and $j$ are related, as follows:
\begin{equation}
  \label{eq:embedding_equation}
  p(i|j)=g(\rho_i^\top\alpha_j)p(i)
\end{equation}

where $p(i)$ is the probability that item $i$ is clicked; $p(i|j)$ is the probability that $i$ is clicked by a user given that $j$ has been clicked by that user. We want the value of $g(.)$ to be large if $i$ and $j$ are frequently clicked by the same users.

There are different choices for the link functions, and an appropriate choice is $g(i,j)=\exp\{\rho_i^\top\alpha_j\}$. Eq. \ref{eq:embedding_equation} can be rewritten as:
\begin{equation}
  \label{eq:embedding_2}
  \log\frac{p(i|j)}{p(i)}=\rho_i^\top\alpha_j
\end{equation}

Note that $\log\frac{p(i|j)}{p(i)}$ is the point-wise mutual information (PMI) \cite{church90} of $i$ and $j$, and we can rewrite Eq. \ref{eq:embedding_2} as:
\begin{equation}
  \label{eq:embedding_3}
  PMI(i,j)=\rho_i^\top\alpha_j
\end{equation}
Empirically, PMI can be estimated using the actual number of observations:
\begin{equation}
\label{eq:empirical_pmi}
\widehat{PMI}(i,j)=\log\frac{\#(i,j)|\mathcal{D}|}{\#(i)\#(j)}
\end{equation}
where $\mathcal{D}$ is the set of all item--item pairs that are observed in the click history of all users, $\#(i)$ is the number of users who clicked $i$, $\#(j)$ is the number of users who clicked $j$, and $\#(i,j)$ is the number of users who clicked both $i$ and $j$.

A practical issue arises here: for item pair $(i, j)$ that is not often clicked by the same user, $PMI(i,j)$ is negative, or if they have never been clicked by the same user, $\#(i,j)=0$ and $PMI(i,j)=-\infty$. However, a negative value of PMI does not necessarily imply that the items are not related. The reason may be because the users who click $i$ may not know about the existence of $j$. A common resolution to this is to replace negative values by zeros to form the PPMI matrix \cite{bullinaria07}. Elements of the PPMI matrix $S$ are defined below:
\begin{equation}
  \label{eq:ppmi_matrix_elements}
  S_{ij}=\max\{\widehat{PMI}(i,j),0\}
\end{equation}
We can see that item embedding vectors $\bm{\rho}$ and item context vectors $\bm{\alpha}$ can be obtained by factorizing PPMI matrix $S$. The factorization can be performed by PMF \cite{salakhutdinov2008a}.

\subsection{Joint Model of Ratings and Clicks}
\label{sec:generative_collaborative_item_embedding_model}
In modeling items, we let item feature vector $\beta_i$ deviate from embedding vector $\rho_i$. This deviation (i.e., $\beta_i-\rho_i$) accounts for the contribution of rating data in the item representation. In detail, if item $i$ has few prior ratings, this deviation should be small; in contrast, if $i$ has many prior ratings, this deviation should be large to allow more information from ratings to be directed toward the item representation. This deviation is introduced by letting $\beta_i$ be a Gaussian distribution with mean $\rho_i$:
\begin{equation}
  \label{eq:prior_of_beta}
  p(\bm{\beta}|\bm{\rho},\sigma_\beta^2) = \prod_i\mathcal{N}(\beta_i|\rho_i,\sigma_\beta^2)
\end{equation}

Below is the \textit{generative process} of the model:
\begin{enumerate}
  \item \textbf{Item embedding model}
  \begin{enumerate}
    \item For each item $i$: draw embedding and context vectors
    \begin{equation}
      \rho_i \propto \mathcal{N}(0, \sigma^2_\rho\mathbf{I}), \quad \alpha_i \propto \mathcal{N}(0, \sigma^2_\alpha\mathbf{I})
    \end{equation}
    \item For each pair $(i, j)$, draw $S_{ij}$ of the PPMI matrix:
    \begin{equation}
      S_{ij} \propto \mathcal{N}({\rho}^\top_i{\alpha}_j,\sigma^{2}_S)
    \end{equation}
  \end{enumerate}
  \item \textbf{Rating model}
  \begin{enumerate}
    \item For each user $u$: draw user feature vector and bias term
    \begin{equation}
      \theta_u \propto \mathcal{N}(0, \sigma^2_\theta\mathbf{I}), \quad b_u \propto \mathcal{N}(0, \sigma^2_b)
    \end{equation}
    \item For each item $i$: draw item feature vector and bias term
    \begin{equation}
      \epsilon_i \propto \mathcal{N}(0, \sigma^2_\epsilon\mathbf{I}), \quad c_i \propto \mathcal{N}(0, \sigma^2_c)
    \end{equation}
    \item For each pair $(u,i)$: draw the rating
    \begin{equation}
      R_{uj} \propto \mathcal{N}(B_{ui}+\theta^\top_u{\beta}_j,\sigma^2_R)
    \end{equation}
  \end{enumerate}
\end{enumerate}

The \textit{posterior distribution} of the model parameters given the rating matrix $R$, PPMI matrix $S$ and the hyper-parameters is as follows.
\begin{equation}
  \label{eq:posterior_joint_model}
  \begin{aligned}
  p(\bm{\Omega}|R,S,\bm{\Theta}) & \propto P(R|\mathbf{b},\mathbf{c},\bm{\theta},\bm{\beta},\sigma_R^2)P(S|\bm{\rho},\bm{\alpha},\sigma_S^2)\\
  & \quad \times p(\bm{\theta}|\sigma_\theta^2)p(\bm{\beta}|\bm{\rho})p(\bm{\rho}|\sigma_\rho^2)p(\bm{\alpha}|\sigma_\alpha^2)p(\mathbf{b}|\sigma_b^2)p(\mathbf{c}|\sigma_c^2)\\
  & = \prod_{(u,i)\in\mathcal{R}}\mathcal{N}(R_{ui}|\theta_u^\top\beta_i,\sigma_S^2)\prod_{(i,j)\in\mathcal{S}}\mathcal{N}(S_{ij}|\rho_i^\top\alpha_j,\sigma_S^2)\\
  & \quad \times \prod_u \mathcal{N}(\theta_u|\mathbf{0},\sigma_\theta^2)\prod_i\mathcal{N}(\beta_i|\rho_i,\sigma_\beta^2)\prod_i\mathcal{N}(\rho_i|\mathbf{0}, \sigma_\rho^2)\\
  & \quad \times \prod_j\mathcal{N}(\alpha_j|\mathbf{0},\sigma_\alpha^2)\prod_u\mathcal{N}(b_u|0,\sigma_b^2)\prod_i\mathcal{N}(c_i|0,\sigma_c^2)
  \end{aligned}
\end{equation}
where $\bm{\Theta}=\{\sigma_\theta^2, \sigma_\beta^2,\sigma_\rho^2,\sigma_\alpha^2,\sigma_b^2,\sigma_c^2\}$.

\subsection{Parameter Learning}
Since learning the full posterior of all model parameters is intractable, we will learn the maximum a posterior (MAP)  estimates. This is equivalent to minimizing the following error function.
\begin{equation}
  \label{eq:objective_function}
  \begin{aligned}
\mathcal{L}(\bm{\Omega})& = \frac{1}{2}\sum_{(u,i)\in \mathcal{R}}[R_{ui}-(B_{ui}+{\theta}^\top_u{\beta}_i)]^2+ \frac{\lambda}{2}\sum_{(i,j)\in \mathcal{S}}(S_{ij}-{\rho}^\top_i{\alpha}_j)^2\\
  & \quad +\frac{\lambda_\theta}{2}\sum_{u=1}^{N}||{\theta}_u||^2_F+\frac{\lambda_\beta}{2}\sum_{i=1}^{M}||\beta_i-\rho_i||^2_F+\frac{\lambda_\rho}{2}\sum_{i=1}^{M}||\rho_i||^2_F\\
  & \quad +\frac{\lambda_\alpha}{2}\sum_{j=1}^{M}||{\alpha}_j||^2_F+\frac{\lambda_b}{2}\sum_{u=1}^Nb_u^2+\frac{\lambda_c}{2}\sum_{i=1}^Mc_i^2
  \end{aligned}
\end{equation}
where $\lambda=\sigma_R^2/\sigma_S^2,\lambda_\theta=\sigma_R^2/\sigma_\theta^2,\lambda_\beta=\sigma_R^2/\sigma_\beta^2,\lambda_\rho=\sigma_R^2/\sigma_\rho^2$, and $\lambda_\alpha=\sigma_R^2/\sigma_\alpha^2$.

We optimize this function by coordinate descent; which alternatively, updates each of the variables $\{\theta_u,\beta_i,\rho_i,\alpha_j,b_u,c_i\}$ while the remaining are fixed.

For $\theta_u$: given the current estimates of the remaining parameters, taking the partial deviation of $\mathcal{L}(\bm{\Omega})$ (Eq. \ref{eq:objective_function}) with respect to $\theta_u$ and setting it to zero, we obtain the update formula:
\begin{equation}
  \label{eq:update_rule_theta}
  \theta_u=\Big(\sum_{i\in \mathcal{R}_u} \beta_i\beta_i^\top+\lambda_\theta\mathbf{I}_d\Big)^{-1}\sum_{i\in \mathcal{R}_u} (R_{ui}-B_{ui})\beta_i
\end{equation}

Similarly, we can obtain the update equations for the remaining parameters:
\begin{equation}
  \label{eq:update_rule_item}
  \beta_i =\Big(\sum_{u\in \mathcal{R}_i} \theta_u\theta_u^\top+\lambda_\beta\mathbf{I}_d\Big)^{-1}\Big[\lambda_\beta\rho_i+\sum_{u\in \mathcal{R}_i}  (R_{ui}-B_{ui})\theta_u\Big]
\end{equation}
\begin{equation}
  b_u=\frac{\sum_{i\in\mathcal{R}_u}R_{ui}-\Big[\mu|\mathcal{R}_u|+\sum_{i\in\mathcal{R}_u}(c_i+\theta_u^\top\beta_i)\Big]}{|\mathcal{R}_u|+\lambda_b}
\end{equation}
\begin{equation}
  c_i=\frac{\sum_{u\in\mathcal{R}_i}R_{ui}-\Big[\mu|\mathcal{R}_i|+\sum_{u\in\mathcal{R}_i}(b_u+\theta_u^\top\beta_i)\Big]}{|\mathcal{R}_i|+\lambda_c}
\end{equation}
\begin{equation}
  \begin{split}
  \rho_i & =\Big[\lambda\sum_{j\in \mathcal{S}_i} \alpha_j \alpha_i^\top+(\lambda_\beta+\lambda_\rho)\mathbf{I}_d\Big]^{-1}\Big(\lambda_\beta\beta_i+\lambda\sum_{j\in \mathcal{S}_i} S_{ij}\alpha_j\Big)
  \end{split}
\end{equation}
\begin{equation}
  \alpha_j  =\Big(\lambda\sum_{i\in \mathcal{S}_j} \rho_i\rho_i^\top+\lambda_\alpha\mathbf{I}_d\Big)^{-1}\Big(\lambda\sum_{i\in \mathcal{S}_j} S_{ij}\rho_i\Big)
\end{equation}

\subsubsection{Computational complexity.} For user vectors, as analyzed in \cite{hu2008collaborative}, the complexity for updating $N$ users in an iteration is $\mathcal{O}(d^2|\mathcal{R}^+|+d^3N)$. For item vector updating, we can also easily show that the running time for updating $M$ items in an iteration is $\mathcal{O}(d^2(|\mathcal{R}^+|+|\mathcal{S}^+|)+d^3M)$. We can see that the computational complexity linearly scales with the number of users and the number of items. Furthermore, this algorithm can easily be parallelized to adapt to large scale data. For example, in updating user vectors $\boldsymbol\uptheta$, the update rule of user $u$ is independent of other users' vectors, therefore, we can update $\theta_u$ in parallel.

\subsection{Rating Prediction}
We consider two cases of rating predictions: \textbf{in-matrix} prediction and \textbf{out-matrix} prediction. In-matrix prediction refers to the case where we predict the rating of user $u$ to item $i$, where $i$ has not been rated by $u$ but has been rated by at least one of the other users; while out-matrix prediction refers to the case where we predict the rating of user $u$ to item $i$, where $i$ has not been rated by any users (i.e., only click data is available for $i$). The missing rating $r_{ui}$ can be predicted using the following formula:
\begin{align}
\hat{r}_{ui} & \approx \mu+b_u+c_i+\theta_u^\top\beta_i
\end{align}

\section{Empirical Study}
\label{sec:empirical_study}
\subsection{Datasets, Competing Methods, Metric and Parameter settings}
\subsubsection{Datasets.}We use three public datasets of different domains with varying sizes. The datasets are: (1) \textit{MovieLens 1M}: a dataset of user-movie ratings, which consists of 1 million ratings in the range $1-5$ to $4000$ movies by $6000$ users, (2) \textit{MovieLens 20M}: another dataset of user-movie ratings, which consists of 20 million ratings in range $1-5$ to 27,000 movies by 138,000 users, and (3) \textit{Bookcrossing}: a dataset for user-book ratings, which consists of 1,149,780 ratings and clicks to 271,379 books by 278,858 users.

Since Movielens datasets contain only rating data, we artificially create the click data and rating data following \cite{DBLP:conf/icdm/BellK07}. Click data is obtained by binarizing the original rating data; while the rating data is obtained by randomly picking with different percentages ($10\%$, $20\%$, $50\%$) from the original rating data. Details of datasets obtained are given in Table \ref{tab:ml_subsets_ml1m} and Table \ref{tab:ml_subsets_ml20m}.
\begin{table}
\parbox{.47\linewidth}{
\centering
\caption{Datasets obtained by picking ratings from the \textit{Movielens 1M}}
\label{tab:ml_subsets_ml1m}
\setlength{\tabcolsep}{0.4em}
\begin{tabular}{c|c|c}
    \hline
    Dataset &\thead{\% rating\\picked} & \thead{Density of\\rating matrix (\%)}\\
    \hline
    ML1-10 & 10\% & 0.3561\\
    ML1-20 & 20\% & 0.6675\\
    ML1-50 & 50\% & 1.6022\\
    \hline
\end{tabular}
}
\hfill
\parbox{.47\linewidth}{
\centering
\caption{Datasets obtained by picking ratings from \textit{Movielens 20M}}
\label{tab:ml_subsets_ml20m}
\setlength{\tabcolsep}{0.4em}
\begin{tabular}{c|c|c}
    \hline
    Dataset &\thead{\% rating\\picked} & \thead{Density of\\rating matrix (\%)}\\
    \hline
    ML20-10 & 10\% & 0.0836\\
    ML20-20 & 20\% & 0.1001\\
    ML20-50 & 50\% & 0.2108\\
    \hline
\end{tabular}
}
\end{table}


From each dataset, we randomly pick 80\% of rating data for training the model, while the remaining 20\% is for testing. From the training set, we randomly pick 10\% as the validation set.

As discussed in Section \ref{sec:methodology}, we consider two rating prediction tasks: \textit{in-matrix} prediction and \textit{out-matrix} prediction. In evaluating the in-matrix prediction, we ensure that all the items in the test set appear in the training set. In evaluating the out-matrix prediction, we ensure that none of the items in the test set appear in the training set.

\subsubsection{Competing methods.} We compare EMB-MF with the following methods.
\begin{itemize}
\item \textit{State-of-the-art methods} in rating predictions: PMF \cite{salakhutdinov2008a}, SVD++ \cite{koren2008factorization}\footnote{The results are obtain by using the LibRec library: http://librec.net/}.
\item \textit{Item2Vec+MF}: The model was obtained by training item embedding and MF separately. First we trained an Item2Vec model \cite{confrecsysBarkanK16} on click data to obtain the item embedding vectors $\rho_i$. We then fixed these item embedding vectors and used them as the item feature vectors $\beta_i$ for rating prediction.
\end{itemize}

\subsubsection{Metric.}
We used Root Mean Square Error (RMSE) to evaluate the accuracy of the models. RMSE measures the deviation between the rating predicted by the model and the true ratings (given by the test set), and is defined as follows.
\begin{equation}
\label{eq:RMSE}
RMSE=\sqrt{\frac{1}{|Test|}\sum_{(u,i)\in Test}(r_{ui}-\hat{r}_{ui})^2}
\end{equation}
where $|Test|$ is the size of the test set.

\subsubsection{Parameter settings.}
In all settings, we set the dimension of the latent space to $d=20$. For PMF and SVD++, Item2Vec+MF, we used a grid search to find the optimal values of the regularization terms that produced the best performance on the validation set. For our proposed method, we explored different settings of hyper-parameters to study the effectiveness of the model.

\subsection{Experimental Results}
The test RMSE results for in-matrix and out-matrix prediction tasks are reported in Table \ref{tb:in_matrix_prediction_subsets} and Table \ref{tb:out_matrix_prediction_subsets}.
\begin{table}[ht]
  \caption{Test RMSE of in-matrix prediction. For EMB-MF, we fixed $\lambda=1$ and used the validation set to find optimal values for the remaining hyper-parameters}
  \label{tb:in_matrix_prediction_subsets}
  \centering
  \setlength{\tabcolsep}{0.2em}
  \begin{tabular}{l|ccc||ccc||c}
    \hline
    \multirow{2}{*}{Methods} & \multicolumn{3}{c||}{ML-1m} & \multicolumn{3}{c||}{ML-20m} & \multirow{2}{*}{Bookcrossing}\\
    & ML1-10 & ML1-20 & ML1-50 & ML20-10 & ML20-20 & ML20-50\\
    \hline
    PMF & 1.1026 & 0.9424 & 0.8983 & 1.0071 & 0.8663 & 0.8441 & 2.1663\\
    SVD++ & 0.9825 & 0.9066 & 0.8871 & 0.8947 & 0.8348 & 0.8191 & 1.6916\\
    Item2Vec+MF & 0.9948 & 0.9135 & 0.8984 & 0.9098 & 0.8527 & 0.8355 & 1.9014\\
    EMB-MF (our) & \textbf{0.9371} & \textbf{0.8719} & \textbf{0.8498}& \textbf{0.8767} & \textbf{0.8299} & \textbf{0.8024} & \textbf{1.6558}\\
    \hline
  \end{tabular}
\end{table}
\begin{table}[ht]
  \caption{Test RMSE of out-matrix prediction. For EMB-MF, we fixed $\lambda=1$; the remaining hyper-parameters are determined using the validation set. Only \textit{Item2Vec} is compared, because PMF and SVD++ cannot be used for out-matrix prediction}
  \label{tb:out_matrix_prediction_subsets}
  \centering
  \setlength{\tabcolsep}{0.2em}
  \begin{tabular}{l|ccc|ccc|c}
    \hline
    \multirow{2}{5em}{Methods} & \multicolumn{3}{c|}{ML-1m} & \multicolumn{3}{c|}{ML-20m} & \multirow{2}{*}{Bookcrossing}\\
    & ML1-10 & ML1-20 & ML1-50 & ML20-10 & ML20-20 & ML20-50\\
    \hline
    Item2Vec+MF & 1.0986 & 1.0365 & 1.039 & 1.0128 & 0.9582 & 0.9784 & 1.7027\\
    EMB-MF (our) & \textbf{1.0312} & \textbf{1.0059} & \textbf{1.0132} & \textbf{0.9729} & \textbf{0.9422} & \textbf{0.9494} & \textbf{1.6828}\\
    \hline
  \end{tabular}
\end{table}

From the experimental results, we can observe that:
\begin{itemize}
  \item The proposed method (EMB-MF) outperforms all competing methods for all datasets on both in-matrix and out-matrix predictions.
  \item EMB-MF, SVD++ and Item2Vec+MF are much better than PMF, which use rating data only. This indicates that exploiting click data is a key factor to increase the prediction accuracy.
  \item For all methods, the accuracies increase with the density of rating data. This is expected because the rating data is reliable for inferring users' preferences.
  \item In all cases, the differences between EMB-MF with the competing methods are most pronounced in the most sparse subsets (ML1-10 or ML20-10). This demonstrates the effectiveness of EMB-MF on extremely sparse data.
  \item EMB-MF outperforms Item2Vec+MF although these two models are based on similar assumptions. This indicates the advantage of training these models jointly, rather than training them independently.
\end{itemize}

\subsubsection{Impact of parameter $\lambda_\beta$.} $\lambda_\beta$ is the parameter that controls the deviation of $\beta_i$ from the item embedding vector $\rho_i$ (see Eq. \ref{eq:prior_of_beta} and Eq. \ref{eq:objective_function}). When $\lambda_\beta$ is small, the value of $\beta_i$ is allowed to diverge from $\rho_i$; in this case, $\beta_i$ mainly comes from rating data. On the other hand, when $\lambda_\beta$ increases, $\beta_i$ becomes closer to $\rho_i$; in this case $\beta_i$ mainly comes from click data. The test RMSE is given in Table \ref{tb:in_matrix_prediction_varying_lambda_beta}.

\begin{table}[htp]
  \caption{Test RMSE of in-matrix prediction task by the proposed method over different values of $\lambda_\beta$ while the remaining hyper-parameters are fixed.}
  \label{tb:in_matrix_prediction_varying_lambda_beta}
  \centering
  \setlength{\tabcolsep}{0.8em}
  \begin{tabular}{c|ccccccc}
    \hline
    $\lambda_\beta$& 0.1 & 1.0 & 10.0 & 20.0 & 50.0 & 100.0 & 1000.0\\
    \hline
    ML1-10 & 1.1301 & 0.9971 & 0.9318 & 0.9381 & 0.9527 & 0.9651 & 0.9924\\
    ML1-20 & 1.1107 & 0.9963 & 0.8911 & 0.8756 & 0.8723 & 0.8769 & 0.8885\\
    ML1-50 & 0.9798 & 0.9193 & 0.8634 & 0.8545 & 0.8512 & 0.8539 & 0.8626\\
    \hline
  \end{tabular}
\end{table}

We can observe that, for small values of $\lambda_\beta$, the model produces low prediction accuracy (high test RMSE). The reason is that when $\lambda_\beta$ is small, the model mostly relies on the rating data which is very sparse and cannot model items well. When $\lambda_\beta$ increases, the model starts using click data for prediction, and the accuracy will increase. However, when $\lambda_\beta$ reaches a certain threshold, the accuracy starts decreasing. This is because when $\lambda_\beta$ is too large, the representations of items mainly come from click data. Therefore, the model becomes less reliable for modeling the ratings.

\section{Related Work}
Exploiting click data for addressing the cold-start problem has also been investigated in the literature. Co-rating \cite{conf/cikm/LiuXZY10} combines explicit (rating) and implicit (click) feedback by treating explicit feedback as a special kind of implicit feedback. The explicit feedback is normalized into the range $[0, 1]$ and is summed with the implicit feedback matrix with a fixed proportion to form a single matrix. This matrix is then factorized to obtain the latent vectors of users and items.

Wang et. al.\cite{conf/pakdd/WangRZW12} proposed Expectation-Maximization Collaborative Filtering (EMCF) which exploits both implicit and explicit feedback for recommendation. For predicting ratings for an item, which does not have any previous ratings, the ratings are inferred from the ratings of its neighbors according to click data.

The main difference between these methods with ours is that they do not have a mechanism for balancing the amounts of click data and rating data when making predictions. In our model, these amounts are controlled depending on the number of previous ratings that the target items have.

Item2Vec \cite{confrecsysBarkanK16} is a neural network based model for learning item embedding vectors using \textit{co-click} information. In \cite{confrecsysLiangACB16}, the authors applied a word embedding technique by factorizing the shifted PPMI matrix \cite{levy2014neural}, to learn item embedding vectors from click data. However, using these vectors directly for rating prediction is not appropriate because click data does not exactly reflect preferences of users. Instead, we combine item embedding with MF in a way that allows rating data to contribute to item representations.

\section{Conclusion}
\label{sec:conclusion}
In this paper, we proposed a probabilistic model that exploits click data for addressing the cold-start problem in rating prediction. The model is a combination of two models: (i) an item embedding model for click data, and (ii) MF for rating prediction. The experimental results showed that our proposed method is effective in rating prediction for items with no previous ratings and also boosts the accuracy of rating prediction for extremely sparse data.

We plan to explore several ways of extending or improving this work. The first direction is to develop a full Bayesian model for inferring the full posterior distribution of model parameters, instead of point estimation which is prone to overfitting. The second direction we are planning to pursue is to develop an online learning algorithm, which updates user and item vectors when new data are collected without retraining the model from the beginning.

\subsubsection*{Acknowledgments.} This work was supported by a JSPS Grant-in-Aid for Scientific Research (B) (15H02789, 15H02703).

\end{document}